\documentclass[]{spie}  
\usepackage[]{graphicx}
\usepackage{amsmath}

\title{Fresnel interferometric arrays for space-based imaging: testbed results} 

\author{Denis Serre\supit{a}, Laurent Koechlin\supit{a}, Paul Deba\supit{a}
\skiplinehalf
\supit{a}Laboratoire d'Astrophysique de Toulouse-Tarbes - Universit\'{e} de Toulouse - CNRS \\14 Avenue Edouard Belin, 31400 Toulouse, France
}

\authorinfo{Authors e-mails can be built using the following syntax: "firstname.lastname@ast.obs-mip.fr"}

\begin{document} 
  \maketitle 

\begin{abstract}
This paper presents the results of a Fresnel Interferometric Array testbed. This new concept of imager involves diffraction focussing by a thin foil, in which many thousands of punched subapertures form a pattern related to a Fresnel zone plate. This kind of array is intended for use in space, as a way to realizing lightweight large apertures for high angular resolution and high dynamic range observations. The chromaticity due to diffraction focussing is corrected by a small diffractive achromatizer placed close to the focal plane of the array.


The laboratory test results presented here are obtained with an 8 centimeter side orthogonal array, yielding a 23 meter focal length at 600 nm wavelength. The primary array and the focal optics have been designed and assembled in our lab. This system forms an achromatic image. Test targets of various shapes, sizes, dynamic ranges and intensities have been imaged. We present the first images, the achieved dynamic range, and the angular resolution.  
\end{abstract}



\section{INTRODUCTION}
\label{sec:intro}
Fresnel arrays are interferometric devices involving many hundreds or thousands of "basic" subapertures: mere holes punched in a large and thin foil. For each of these individual subapertures, diffraction occurs. 
The subapertures positioning law is designed so that the resulting interference produces the desired wavefront shape: for instance a spherical wavefront.
The circular Fresnel Zone Plate (also called Soret zone plate), constituted of alternatively opaque and transparent concentric rings, is an example of Fresnel array: the distance from one transmissive ring to the other is adjusted so that the optical path to a point named "focus" differs of one $\lambda$ from one ring to the other. As a result constructive interference occurs at this point, resulting in an equivalent of focalisation: concentration of light.

In our approach, the orthogonal geometry has been preferred to Soret's concentric rings for two main reasons: its allowance of vacuum rather than a transparent material for transmissive zones, and the higher dynamic range allowed in most of the image field. This optical scheme strongly reduces manufacturing and positioning constraints compared to those of standard interferometric arrays and monolithic mirrors. It also yields high dynamic range as well as larger field-resolution ratios than classical imaging interferometers. We propose this concept for high angular resolution imaging in future space projects (Koechlin et al 2005 \cite{Koechlin_aa_2005}) .

\begin{figure}
\begin{center}
\begin{tabular}{c}
\includegraphics[width=0.7\linewidth]{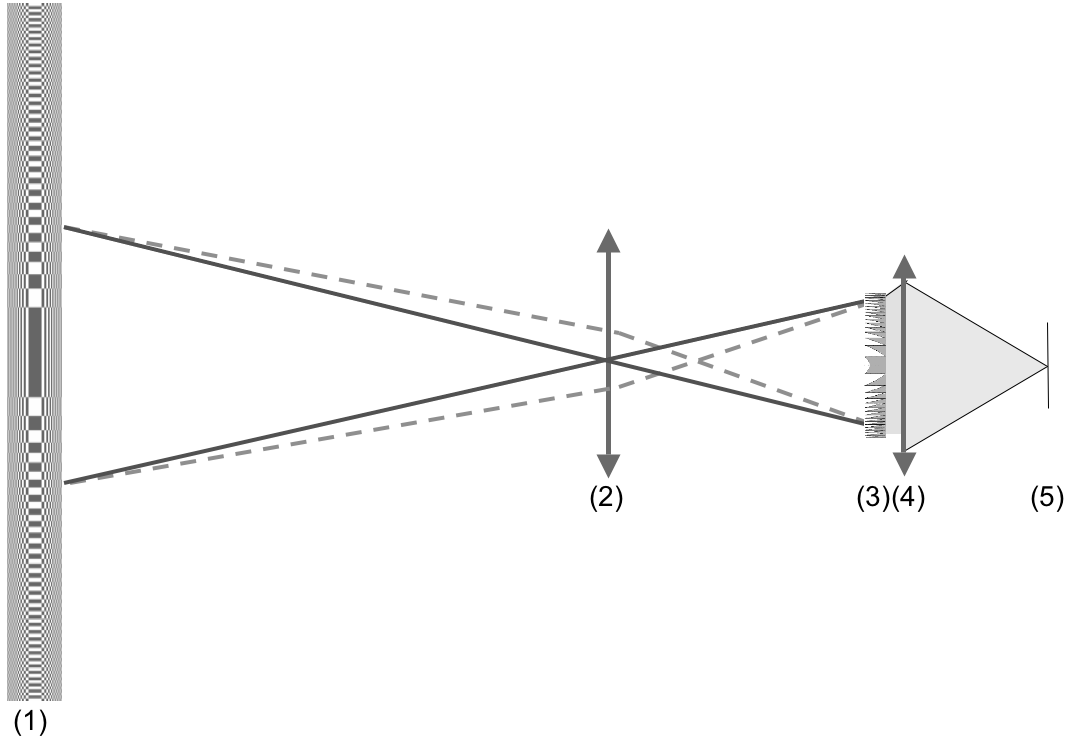}
\end{tabular}
\end{center}
\caption[Achromatisation scheme] 
{ \label{fig:achromatisation_scheme}
An orthogonal Fresnel array is placed on plane (1) and used at its first order of interference. Different wavelengths (dashed and dotted lines) are focussed at different distances from the Fresnel array. A field optics (2) images the Fresnel array onto a pupil plane (3), where a diverging Fresnel zone lens is placed. The emerging beam is perfectly achromatic, but needs a classical re-imaging optics to be convergent (4). The final achromatic image is formed on plane (5).}
\end{figure}

An orthogonal Fresnel array of side $c$ involving $N$ Fresnel zones results in a focal length $f$ at its first order of interference, which can be written \cite{Koechlin_aa_2005}:
\begin{equation}
\centerline{$f = \frac{c^2}{8N}\,\frac{1}{\lambda}$ .}
\label{eq:lambda_f}
\end{equation}
The unavoidable chromaticity of this kind of focussing element can be corrected using an optical scheme based on Schupmann studies \cite{Schupmann_1899} , and emerging from the chromatic correction of holograms developed by Faklis \& Morris in 1989 \cite{Faklis_oe_1989} . The principle is to place in a pupil plane an optical device whose chromatic dispersion is opposite to that of the Fresnel array. That means in our case to use a diffractive optical device used at the order -1: a diverging Fresnel zone lens whose focal length varies in $-\frac{1}{\lambda}$. Fig.\ref{fig:achromatisation_scheme} presents the optical scheme.

Following [Koechlin et al 2004 \cite{Koechlin_spie_2004}] and [Serre et al 2005 \cite{Serre_jithd_2006}] , and thanks to a CNES{\footnote {Centre National d'Etudes Spatiales
}} R\&T funding, we developed in the past two years at Universit\'e Paul Sabatier a Fresnel Interferometric Array breadboard testbed. The main focussing element is an 8 cm side square array and it is associated to optimized focal optics, leading to achromatic images and allowing diffraction limited performances. We present in this paper: firstly the constitutive elements of this prototype, and secondly the measured performances.

\section{CONSTITUTIVE ELEMENTS}

\subsection{Fresnel interferometric array}
\begin{figure}
\begin{minipage}[c]{0.5\linewidth}
\begin{center}
\includegraphics[width=0.9\linewidth]{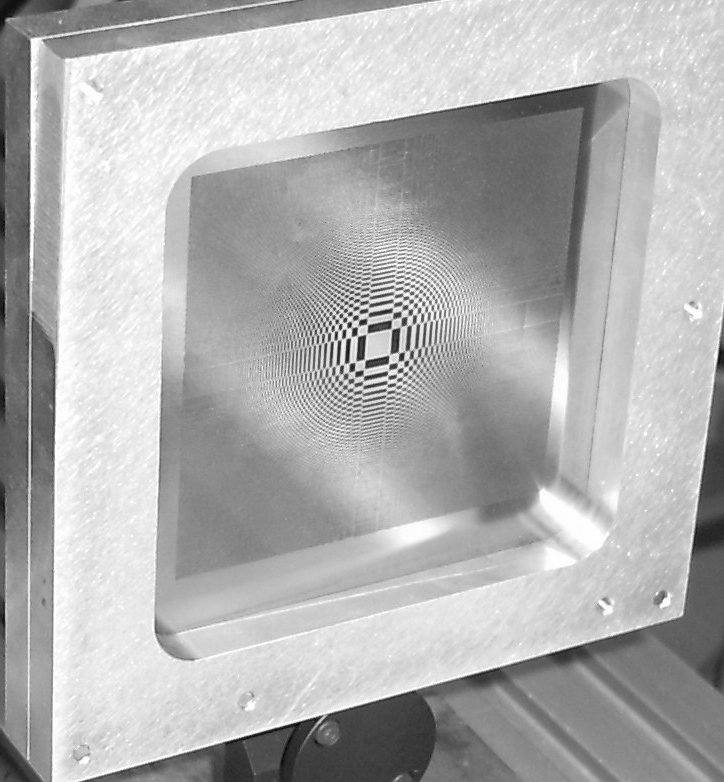}
\newline
\end{center}

\end{minipage}
\hspace*{0.5cm}
\begin{minipage}[c]{0.4\linewidth}

\caption[Carved Fresnel interferometric array]
 {26680 holes are carved on this 8x8 cm$^2$ surface. The focal length is 23m for $\lambda = 600$nm. The surface is not perfectly plane due to the manufacturing process, but this has no measurable consequence on the image quality thanks to the high manufacturing tolerance of this kind of optics, as the results will show.}
\includegraphics[width=0.9\linewidth]{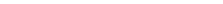}
\label{fig:interferometric_array}

\end{minipage}
\end{figure}

Our Fresnel interferometric array is an 8 cm side orthogonal array, 58 Fresnel zones, i.e. 26680 subapertures, carved by UV laser in an 80 microns thick stainless steal foil. The resulting focal length is 23m at $\lambda = 600$nm. In order to allow mechanical resistance of the array, the size of the individual apertures has been slightly decreased compared to the maximal possible dimensions. In our case, the smallest (outer) apertures are 140x140 microns squares instead of 170*170. This focussing array can be seen on fig.\ref{fig:interferometric_array}.

\subsection{Focal optics}
\begin{figure}
\begin{center}
\includegraphics[width=0.9\linewidth]{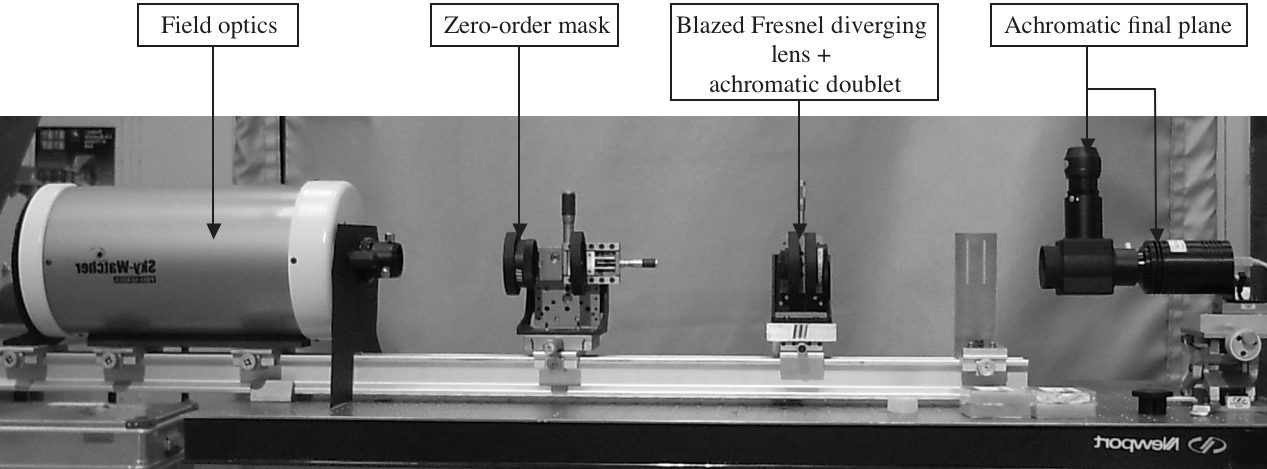}
\end{center}

\caption[Focal optics]
 {The breadboard design focal optics directly emerges from the achromatisation principle presented in fig.\ref{fig:achromatisation_scheme}: field optics (a Maksutov telescope with off axis diaphragm), the diverging Fresnel zone lens, and an image reformation optical element are present. A mask is placed at the Maksutov focus, allowing the elimination of the zero-order.}
\label{fig:achromatisation_optics}

\end{figure}

\begin{figure}
\begin{minipage}[c]{0.5\linewidth}
\begin{center}
\includegraphics[width=0.9\linewidth]{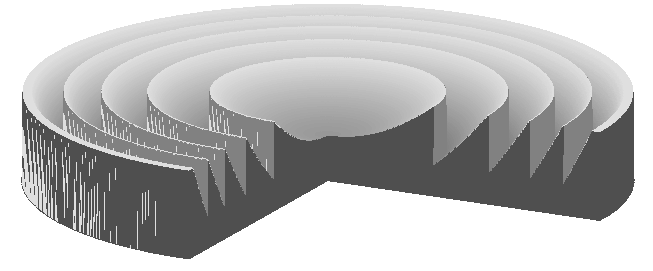}
\newline
\end{center}

\end{minipage}
\hspace*{0.5cm}
\begin{minipage}[c]{0.4\linewidth}

\caption[Blazed diverging Fresnel zone lens]
 {Representation of the five central zones of the used Fresnel blazed diverging lens. This lens has been carved by ionic engraving on a fused silica flat window (5 mm thickness). The depth of all grooves is $\simeq$1.37 microns, this value being nominal for a 600 nm wavelength and the refraction index of silica. The width of the grooves decreases from 600 microns (central zone) to 35 microns (most external pattern).}
\includegraphics[width=0.9\linewidth]{figs/ligne_blanche.png}
\label{fig:profil_blazee_5N_3D}

\end{minipage}
\end{figure}

\begin{figure}
\begin{minipage}[c]{0.5\linewidth}
\begin{center}
\includegraphics[width=0.9\linewidth]{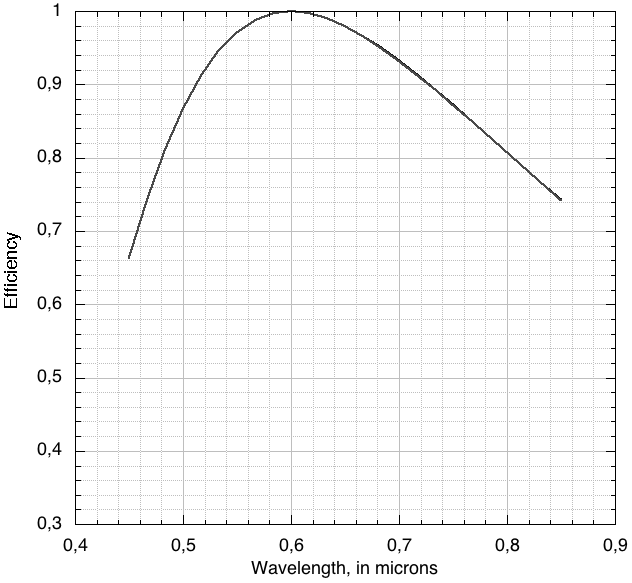}
\newline
\end{center}

\end{minipage}
\hspace*{0.5cm}
\begin{minipage}[c]{0.4\linewidth}

\caption[Blazed diverging Fresnel zone lens efficiency]
 {The efficiency of the lens is defined as the quantity of energy in the central peak of the Fresnel zone lens PSF, compared to the energy which would be obtained if the emerging wavefront was perfectly spherical.\\
Here are presented the dependency of efficiency on wavelength, in two cases: lens manufactured with continuous slopes (nominal configuration, not our case), and lens manufactured with slopes approximated by a finite number of steps (our case, 128 steps). The two curves overlap, the resulting efficiencies being very close. The maximal efficiency is for $\lambda = 600$nm, which is the wavelength the lens has been designed for, but remains better than 90\% through a 200nm wide spectral bandwidth.
  }
\includegraphics[width=0.9\linewidth]{figs/ligne_blanche.png}
\label{fig:blazed_fresnel_efficiency}

\end{minipage}
\end{figure}

The focal optics are the direct application of the chromatic correction scheme presented in fig.\ref{fig:achromatisation_scheme}. The elements are (see fig.\ref{fig:achromatisation_optics}):\\
- a field lens: in fact a Maksutov telescope associated to a 37mm diameter diaphragm, used off-axis to avoid central obturation;\\
- pupil optics: a 116 zones diverging Fresnel lens, carved by ionic engraving using successive masks on a fused silica flat window. The useful diameter is 16 mm, which corresponds to the diagonal of the orthogonal Fresnel array imaged by the field optics. The five central zones of this lens can be seen for illustration on fig.\ref{fig:profil_blazee_5N_3D}, and its efficiency dependance with wavelength can be seen on fig.\ref{fig:blazed_fresnel_efficiency}. The profile of this lens and the efficiency dependance have been obtained following the method exposed in [Serre 2007 \cite{Serre_ea_2007}] ({\it submitted}). The individual grooves are 1.37 microns in depth, their different profiles are discretized with 128 depth levels, allowing transmission into order $-1$ better than 90\% through a $\frac{\Delta \lambda}{\lambda} = 0.3$ spectral bandwidth.\\
- an achromatic doublet allows the formation of a real (i.e. non virtual) image on a B\&W CCD camera. An eyepiece can also be placed for direct control.\\
A narrow cross mask allowing the elimination of order zero of the Fresnel array is placed at the focal plane of the Maksutov telescope.

\subsection{Source simulation}
As the breadboard testbed is situated in a clean room, we need artificial sources. They are not really constitutive elements of the Fresnel interferometric imager, but are necessary to evaluate it. We disposed different types of sources in the focal plane of a parabolic mirror, collimating them along the optical axis of the Fresnel array. These different types can be source points, uniform discs or more shaped targets (see "galaxy" image in next section). LEDs or halogene illumination, with or without spectral filters, has been used.

\section{Measurements}

\subsection{Angular resolution and chromatic correction}
\begin{figure}[!htbp]
\begin{minipage}[c]{0.45\linewidth}
\begin{center}
\includegraphics[width=0.9\linewidth]{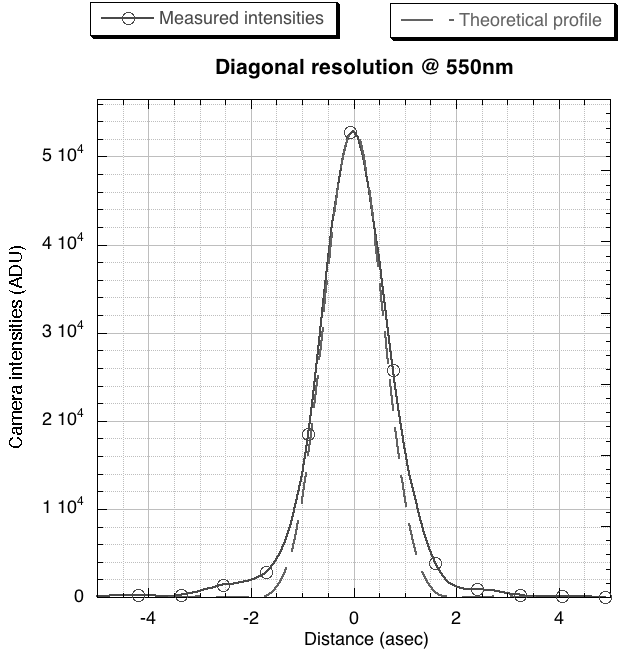}
\end{center}
\end{minipage}
\hspace*{0.5cm}
\begin{minipage}[c]{0.45\linewidth}
\begin{center}
\includegraphics[width=0.9\linewidth]{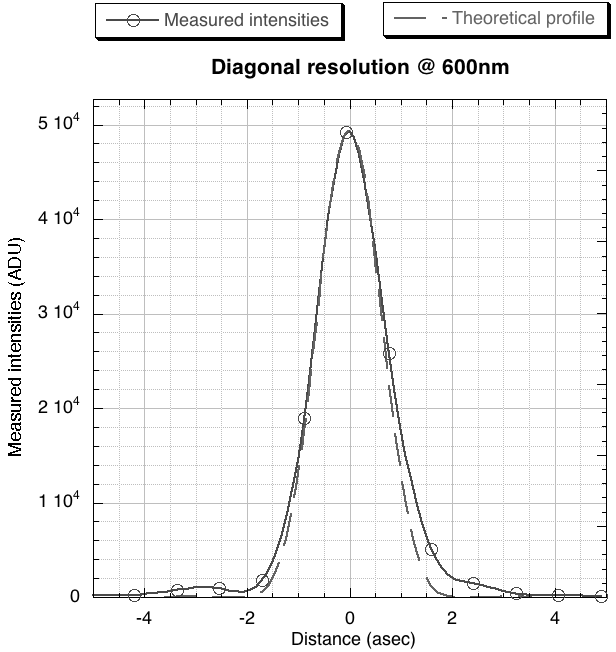}
\end{center}
\end{minipage}

\begin{minipage}[c]{0.45\linewidth}
\begin{center}
\includegraphics[width=0.9\linewidth]{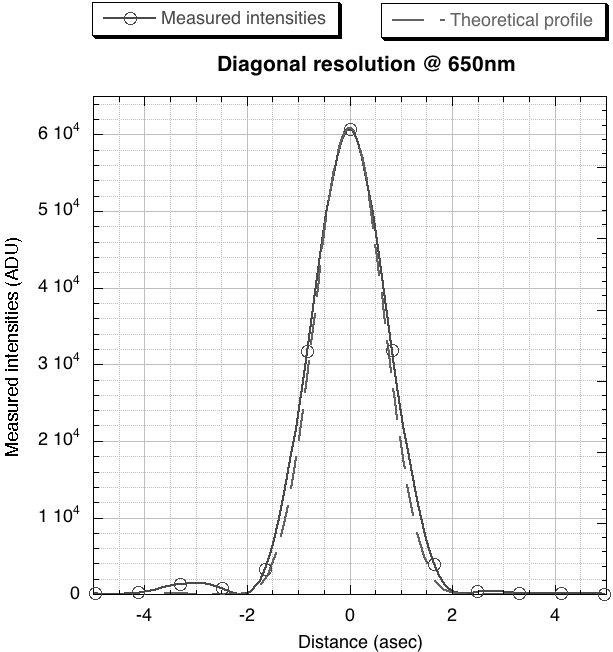}
\end{center}
\end{minipage}
\hspace*{0.5cm}
\begin{minipage}[c]{0.45\linewidth}
\begin{center}
\includegraphics[width=0.9\linewidth]{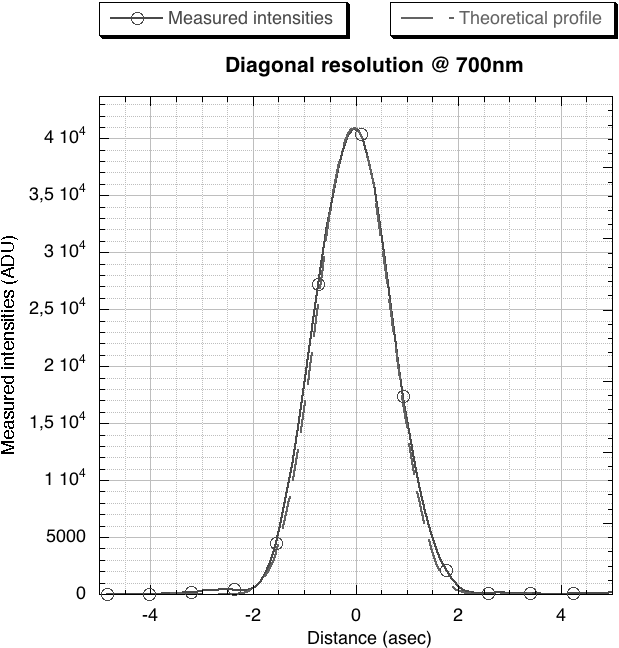}
\end{center}
\end{minipage}

\caption[Angular resolution]
{Two profiles are compared for each of the four wavelengths presented: the theoretical profile, limited by the resolution of the Fresnel interferometric array (dashed profile), and the measured profile on the camera (continuous profile + measurements points: pixels of the camera). Limited by diffraction performances are reached in a wide spectral bandpass, confirming the efficiency of the chromatic correction scheme and the blazed diverging Fresnel zone lens design. Scattered light near central peak may be attributed to turbulence present in the clean room, the effect being more important for short wavelengths.}
\label{fig:angular_resolution_measures}

\end{figure}

The theoretical angular resolution of the whole testbed is $\frac{\lambda}{c} = 1.55$asec (at $\lambda = 600$nm). In order to evaluate the effective resolution, we have used a pinhole (0.5 asec diameter) as a source, illuminated by halogen light and spectral filters centered on 550, 600, 650 and 700 nm (50nm bandpass for each filter). On fig.\ref{fig:angular_resolution_measures} can be seen the comparisons between the theoretical, diffraction limited profiles and the measured profiles. The resolution is limited by the size of the Fresnel array and its associated diffraction limit for these wavelengths. The chromatic correction principle and its realization is consequently confirmed, as well as the design of the blazed diverging Fresnel zone lens.

\subsection{Imaging capabilities}
\begin{figure}
\begin{minipage}[c]{0.5\linewidth}
\begin{center}
\includegraphics[width=0.9\linewidth]{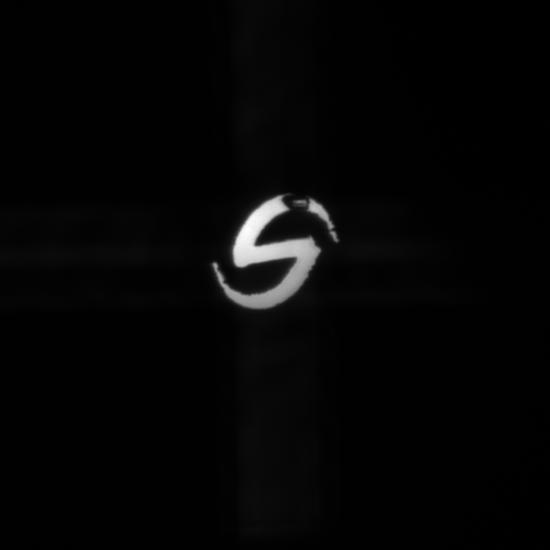}
\end{center}

\end{minipage}
\hspace*{0.5cm}
\begin{minipage}[c]{0.4\linewidth}

\caption[Imaged "galaxy"]
 {The dimension of this "galaxy" target, seen from the Fresnel interferometric array, is 72 asec. The target is illuminated with an halogen source (no spectral filter), illustrating the efficiency of the achromatism scheme and the imaging capabilities of the whole concept. A manufacturing defect can be seen in one of the arms, as well as the faint convoluted diffraction spikes associated to the orthogonal design of the Fresnel array.}
\includegraphics[width=0.9\linewidth]{figs/ligne_blanche.png}
\label{fig:galaxy_blazee}

\end{minipage}
\end{figure}

One can see on fig.\ref{fig:galaxy_blazee} an image obtained using the breadbord testbed: the source is a "galaxy" target carved in a stainless steal foil. The "galaxy" dimension from limb to limb is 450 microns, that is 72 arc seconds as seen from the Fresnel array. The illuminating source is a halogen lamp, whose spectrum spans throughout the visible domain, illustrating the excellent chromatic correction as exposed in previous section.

\subsection{PSF modeling and rejection rate results}
\begin{figure}
\begin{center}
\includegraphics[width=0.9\linewidth]{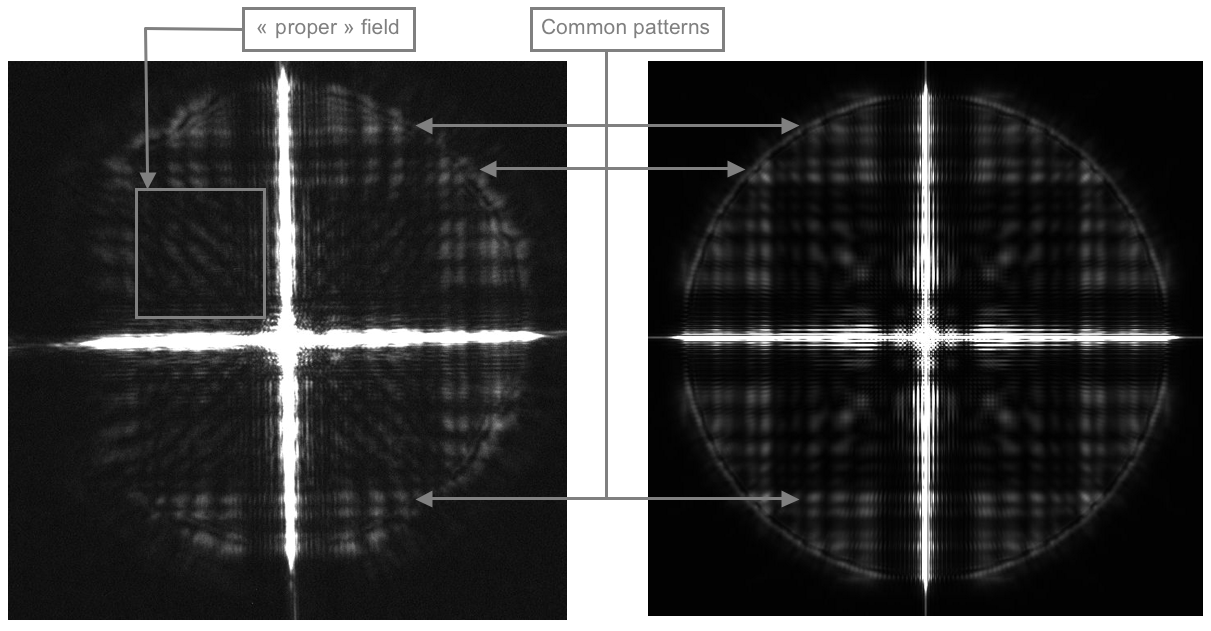}
\end{center}

\caption[Comparison of simulated and obtained PSFs]
 {The left PSF has been obtained using our testbed: a LED with maximal emission at 601 nm illuminating a pinhole. The PSF on the right is a computer simulation taking into account all the optical elements of the testbed, including the stepped groove profiles of the chromatic corrector lens and the spectrum of the source. The two images are saturated in the same way in order to evidence the faint illumination regions. Common patterns can be seen on the two images, and rejection rates are also comparable.}
\label{fig:real_psf_vs_simulated}

\end{figure}

We have developed for the Fresnel interferometric array concept a plane-to-plane propagation software based on Fresnel diffraction algorithms, and have applied it to our breadboard testbed. The algorithms take into account the main orthogonal array, achromatic field optics, a "zero-order mask" placed at the focal point of a parallel beam focussed by the field optics, the diverging Fresnel zone lens with all its groove profiles and the silica index variation, and finally the image reformation achromatic optics. An actual (acquired) PSF can be seen and compared to a simulated one on fig.\ref{fig:real_psf_vs_simulated}. To show the faint patterns of the PSF, the acquired image is highly overexposed.\\
In order to evaluate the rejection rate of the PSF, two images are presently necessary: a first one is acquired to measure the unsaturated level of the central peak, and then a second one is acquired with an exposure time multiplied 1000x times. Defining the rejection rate as the mean level of the "clean" field of the PSF compared to the maximum level of the central peak, we can compare theoretical and measured values. They are of the same order of magnitude (see tab.\ref{tab:rejection_measures}).

\begin{table}[!htbp]
\centering
\begin{tabular}{|c|c|c|}
\hline
Led's maximal & Measured rejection rate  & Theoretical rejection rate\\
intensity wavelength & & \\
\hline
525 nm & 4.6e-6 & 1.2e-6\\
601 nm & 2.3e-6 & 1.2e-6\\
630 nm & 1.9e-6 & 1.5e-6\\
\hline
\end{tabular}
\caption {Comparison of simulated and measured rejection rates. The values are in correct agreement. Turbulence may be invoked to explain why rejection rate comparisons become worst as wavelength decreases.}
\label{tab:rejection_measures}
\end{table}

\section{conclusion}
We have presented in this paper the first results of an orthogonal Fresnel interferometric array breadboard testbed. We are able to confirm the efficiency of the achromatisation scheme retained, as well as the diffraction-limited performances and the "wide" field imaging capabilities. The measured rejection rates are also in good agreement with computer generated previsions.\\
The next steps using this testbed are to evaluate rejection rates with two sources of different luminosity in the same field, to test different manufacturing processes for the blazed diverging Fresnel zone lens, and to assess Fresnel arrays subaperture shapes and layout, allowing higher transmission rates and dynamic ranges.\\
In the next few years, on one hand members of our team (and we hope new people !) aim to evaluate a 2$^{\text{nd}}$ generation of ground-based testbed: its main evolutions will be an increase in dimensions (typically 20 cm side for the Fresnel array), higher number of Fresnel zones, and last but not least: real sky observation. On the other hand, studies will keep on being done to evaluate the performances of this kind of instrument and the potential astrophysical targets of a space-based Fresnel interferometric array concept. This space observatory could have a front array dimension of a few meters or more, and acquire data in the UV, visible or/and IR domains.\\

\footnotesize{This work was supported by Universit\'{e} Paul Sabatier Toulouse III, CNRS, CNES, Thal\`{e}s Alenia Space and the Fonds Social Europ\'{e}en.}

 


\bibliography{spie_07_07_18}   
\bibliographystyle{spiebib}   

\end{document}